\documentclass{jpp}
\usepackage{graphicx}

\usepackage[utf8]{inputenc}
\usepackage[dvipsnames]{xcolor}
\usepackage[T1]{fontenc}
\usepackage{amsmath}
\usepackage[colorlinks=true,
            linkcolor=blue,
            citecolor=blue]
           {hyperref}

\usepackage{cleveref}

\crefformat{section}{\S#2#1#3} 
\crefformat{subsection}{\S#2#1#3}
\crefformat{subsubsection}{\S#2#1#3}

\newcommand{\gradperp}{\nabla_\perp}
\newcommand{\pd}[2]{\frac{\partial #1}{\partial #2}}
\newcommand{\ve}[0]{\mathbf{v}_E}
\newcommand{\vc}[0]{\mathbf{v}_C}

\newcommand{\hermes}[0]{\texttt{hermes-3}}

\shorttitle{Drift-reduced modeling}
\shortauthor{E. Tocco, B. Dudson., I. Abel, B. Zhu}

\title{Drift-reduced fluid modeling of rapidly rotating plasmas}

\author{Edward A. Tocco\aff{1}
  \corresp{\email{eatocco@umd.edu}},
  B. D. Dudson\aff{2},
  I. G. Abel\aff{3} \and B. Zhu\aff{4,5}}

\affiliation{\aff{1}Space Power and Propulsion Laboratory, University of Maryland, College Park, MD 20742, USA
\aff{2}Lawrence Livermore National Laboratory, Livermore, CA 94550, USA
\aff{3}Institute for Research in Electronics and Applied Physics, University of Maryland, College
Park, MD 20742, USA
\aff{4}Department of Applied Physics and Applied Mathematics, Columbia University, New York, NY 10027, USA
\aff{5}Columbia Fusion Research Center, Columbia University, New York, NY 10027, USA}

\begin{document}

\maketitle

\begin{abstract}

In this paper, we investigate the effects of rapid rotation (Mach number $M \gtrsim 1$) on plasma fluid stability, focusing specifically on Kelvin-Helmholtz (KH) and interchange instabilities - including both magnetic-curvature-driven (CDI) and rotation-driven (RDI) interchanges. Building on previous studies of shear flow stabilization, we utilize a drift-reduced fluid approach rather than standard magnetohydrodynamics to capture finite Larmor-radius effects. To achieve this, the drift-reduced equations were modified to include the centrifugal force and implemented in \hermes{} \citep{Dudson2024}, an extension to the \texttt{BOUT++} \citep{Dudson2009} framework. Because plasma rotation both drives the RDI and provides stabilizing shear flow, we find that the global plasma stability is sensitive to background profile characteristics. We identify two distinct regimes of RDI behavior and establish a simple criterion based on the density and velocity profiles to predict RDI susceptibility. This approach is similar to recent local gyrokinetic studies of shear flow that compared instability growth rates to shearing rates \citep{Ivanov2025}. Finally, by examining cases where the plasma is both interchange- and KH-unstable, we find that global KH modes make the plasma less resistant to RDI.

\end{abstract}

\section{Introduction}

The stability of rapidly-rotating plasmas in the long wavelength (as compared to the Larmor-radius) regime has long been of interest, particularly in the study of the centrifugally confined magnetic mirror. These systems are characterized by large inertial forces that can drive interchange-type instabilities, as well as large shear flows that can stabilize the interchange but are themselves potentially unstable to Kelvin-Helmholtz (KH) modes. Given the inherent dependence on the equilibrium profiles, we conduct a global stability analysis of this system. We examine two types of interchange instabilities, one driven by magnetic curvature and one driven by the centrifugal force resulting from rapid azimuthal rotation. We will refer to these as curvature-driven interchange (CDI) and rotation-driven interchange (RDI), respectively.

Building on previous work \citep{Hassam1992, Huang2001,Huang2004,Ng2005}, several key findings emerge. First, analytical models and magnetohydrodynamics (MHD) simulations have demonstrated that interchange instabilities can be stabilized nonlinearly by flow shear, provided that the local shear flow exceeds a threshold determined by the instability's growth rate and a finite viscosity \citep{Hassam1992, Huang2001}. Second, the stabilization of the RDI variant was shown to be incomplete, necessitating particle sources to maintain equilibrium \citep{Huang2004}. Finally, previous studies concluded that finite Larmor-radius (FLR) effects can aid in stabilizing interchange modes via gyroviscosity, although this mechanism was not explored numerically \citep{Ng2005}.

We expand upon this research by incorporating FLR effects into our numerical simulations and exploring the RDI instability in greater depth, as it is highly relevant to a rotating magnetic mirror. We assess the qualitative similarities between these new simulations and prior studies, detailing new insights derived from higher-fidelity modeling. Greater emphasis is also placed on the complex interplay in the RDI between the flow shear and the destabilizing force, both of which originate from the azimuthal $\mathbf{E}\times\mathbf{B}$ velocity.

In this study, we utilize the drift-reduced fluid equations \citep{Simakov2003} implemented in the \texttt{hermes-3} code \citep{Dudson2024}, which is built on the \texttt{BOUT++} framework \citep{Dudson2009}. These are outlined in \cref{sec:drift_reduced}. Drift-reduced models have been used successfully to model the edge region in tokamak plasmas \citep{Dudson2025}. The parameter regime under which these equations were derived corresponds to the parameter regime of interest for this study. We perform simulations in the azimuthal $R-\phi$ plane, which isolates the most unstable $k_\parallel = 0$, or wavelength parallel to the magnetic field, modes. In \cref{sec:linear_instabilities}, we begin by analyzing the KH (\cref{sec:kh}) and interchange (\cref{sec:interchange}) instabilities separately to verify that the model accurately captures the relevant physics. We reproduce linear growth rates for KH and CDI modes and observe FLR effects (\cref{sec:FLR}) in the interchanges. We then introduce flow shear in \cref{sec:shear} to validate the shear stabilization criterion for CDI and demonstrate the stabilizing influence of gyroviscosity (\cref{sec:shear_stablization}). To properly model RDI, we modify the implementation of the vorticity equation by relaxing the often-used Boussinesq approximation \citep{Oberbeck1879, Ross2018}, which otherwise treats the density as a constant in the vorticity evolution equation. While this physics system has been studied previously \citep{Popovich2010}, we extend the study into fully nonlinear, shear-rotational cases with FLR effects. The RDI instability is then analyzed for peaked density and velocity profiles, and a heuristic for stability is proposed and tested in \cref{sec:type2}. 
Finally, in \cref{sec:global}, we present a global stability analysis of profiles that can simultaneously support both KH and interchange modes.

\section{Drift-reduced fluid equations \label{sec:drift_reduced}}

Drift-reduced fluid equations are valid in the low-$\beta$, strongly magnetized regime. While the fluid velocity is generally assumed to be smaller than the sound speed, we relax this assumption by incorporating a supersonic yet slowly-varying background flow. 

Although the drift-reduced equations implemented in \texttt{hermes-3} can accommodate multiple ion species and coupling to a neutral gas via atomic reactions, the simulations presented here utilize a single ion species in the electrostatic limit. Therefore, the model is the same as used in the TCV-X21 verification study \citep{Dudson2025}. Specifically, we solve evolution equations for electron density, ion and electron pressure, and ion vorticity in the azimuthal plane. 

The continuity equation for the number density $n$ is given by
\begin{equation}
    \pd{n}{t} = -\nabla \cdot (n \ve + n\mathbf{v}_{mag,e}) + S_n,
\end{equation}

\noindent where the $\mathbf{E}\times\mathbf{B}$ velocity is 
\begin{equation}
    \mathbf{v}_E = \frac{\mathbf{b}\times \nabla \varphi}{B},\label{eq:ve}
\end{equation}

\noindent and the electron flow due to magnetic drifts is given by  
\begin{equation}
    \mathbf{v}_{mag,e} = -T_e\nabla \times\frac{\mathbf{b}}{B}.
\end{equation}

The ion and electron pressure evolution equations are 
\begin{equation}
    \frac{3}{2}\pd{p_i}{t} = -\nabla \cdot\left(\frac{3}{2}p_i\ve + \frac{5}{2} p_i \mathbf{v}_{mag,i}\right) - p_i \nabla \cdot \ve+ S_{pi} \label{eq:ion_energy},
    \end{equation} and
    \begin{equation}
     \frac{3}{2}\pd{p_e}{t} = -\nabla \cdot\left(\frac{3}{2}p_e\ve + \frac{5}{2} p_e \mathbf{v}_{mag,e}\right) - p_e \nabla \cdot \ve+ S_{pe} \label{eq:electron_energy}, 
\end{equation} respectively.

\noindent Sources for each equation ($S_n, S_{pi},$ and $S_{pe}$) are included as needed to maintain a background profile. 

The vorticity $\omega$ is derived from the divergence of the ion polarization current, and is implemented in \hermes{} as
\begin{equation}
  \omega = \nabla\cdot\left[\frac{m_in_0}{B^2}\nabla_\perp\left(\phi + \frac{p_i}{n_0}\right)\right], \label{eq:vorticity_definition}
\end{equation}
where $\nabla_\perp \equiv \nabla - \mathbf{b}\mathbf{b}\cdot\nabla$ and the Oberbeck-Boussinesq approximation~\citep{Oberbeck1879} is made, replacing the density in the polarisation current with a constant $n_0$. The evolution of the vorticity is derived from current continuity, such that the divergence of the sum of all currents vanishes, in this case polarization current and diamagnetic current $\mathbf{J}_d$.
\begin{eqnarray}
  \frac{\partial \omega}{\partial t} &=& -\nabla\cdot\left[\frac{m_i}{2B^2}\nabla_\perp\left(\mathbf{v}_E \cdot\nabla p_i\right) + \frac{\omega}{2}\mathbf{v}_E + \frac{m_in_0}{2B^2}\nabla_\perp^2\phi\left(\mathbf{v}_E + \frac{\mathbf{b}}{n_0B}\times\nabla p_i\right)\right] \nonumber \\
  &&+ \nabla\cdot\mathbf{J}_d, \label{eq:vorticity}
\end{eqnarray}
\noindent where the divergence of the diamagnetic current is:
\begin{equation}
  \nabla\cdot\mathbf{J}_d = \nabla\cdot\left[\left(p_e + p_i\right)\nabla\times\frac{\mathbf{b}}{B}\right].
  \label{eq:diamagnetic}
\end{equation}

Note that the vorticity equation \eqref{eq:vorticity} has been modified from its original form in \cite{Simakov2003} to improve energy conservation within the model, in addition to applying the Boussinesq approximation. The Boussinesq approximation will be removed in \cref{sec:type2} in order to accurately capture the physics that drives the RDI. We also recast the vorticity equation to explicitly show the inertial term responsible for RDI, with the modifications detailed in \cref{appB}. 

In \hermes{}, for magnetic field $B$ and ion mass $m_i$, lengths are normalized to the ion Larmor radius $L_0 = \rho_i = (1/B)\sqrt{m_i T_0/e}$, and time is normalized to the ion gyrofrequency $t_0 = \Omega_i^{-1} = m_i/eB$. For the remainder of this paper, with the exception of lengths, will be shown in normalized units. 

\subsection{Numerical implementation: \texttt{BOUT++} and \texttt{hermes-3}}

While exhaustive details of the numerical model can be found in \cite{Dudson2024,Dudson2009}, we briefly summarize the key features here. \texttt{BOUT++} is a modular fluid code framework primarily used for the simulation of plasma fluid equations in the tokamak edge. It implements a wide variety of numerical discretization schemes, including both implicit and explicit temporal discretization, and finite difference and Fourier transform methods for spatial discretization. Space can be discretized on an arbitrary curvilinear grid, which is especially useful for modeling the tokamak edge and divertor regions. The equations to be solved are specified in a modular format designed for flexibility and readability. Built upon \texttt{BOUT++}, the \texttt{hermes-3} package serves as a specialized physics module, providing a robust, pre-configured implementation of the drift-reduced fluid equations tailored for magnetized plasma environments. Simulations were performed on one CPU node of the Perlmutter supercomputer.

\section{Linear instabilities \label{sec:linear_instabilities}}

In this section, we verify that the \hermes{} implementation accurately captures both KH and interchange instabilities, including first-order FLR effects. Initial simulations are performed in a 2D slab geometry. In the slab geometry, $x$ corresponds to the radial coordinate (the direction of the density and flow gradients), and $z$ corresponds to the azimuthal coordinate. Later simulations will move to cylindrical coordinates to study the RDI.

\subsection{The Kelvin-Helmholtz instability \label{sec:kh}}
Kelvin-Helmholtz instability arises in sheared flows due to the advection of vorticity perturbations into regions of different velocity. In particular, sheared flows with inflection points are known to be unstable \citep{Rayleigh1879}. In the drift-reduced equations, the term responsible for this instability originates in the polarization current. In the limit of low ion temperature and a constant background magnetic field perpendicular to the plane, \eqref{eq:vorticity} reduces to 
\begin{equation}
    \pd{\omega}{t} = -\nabla \cdot \left(\omega \textbf{v}_E\right), \label{eq:ev_vorticity_simplified}
\end{equation}

\noindent and the expression for the vorticity simplifies to
\begin{eqnarray}
    \omega = \nabla \cdot \left(\frac{m_i n_0}{B^2} \gradperp \phi \right).\label{eq:vorticity_simplified}
\end{eqnarray}

\begin{figure}
    \centering
    \includegraphics[width=\linewidth]{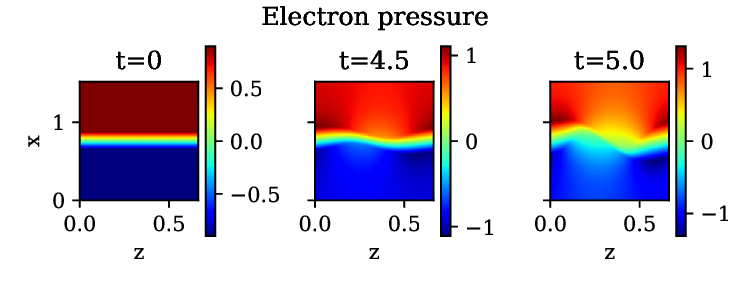}
    \caption{Time slices of velocity for the ramped case, labeled in time units normalized to the ion cyclotron frequency. As the simulation evolves, the sheared profile breaks up and begins to form vortices typical of 2D fluid turbulence.}
    \label{fig:kh_contours}
\end{figure}

With Eq. \eqref{eq:ve} for $\mathbf{v}_E$, \eqref{eq:ev_vorticity_simplified} and \eqref{eq:vorticity_simplified} form a closed system. If $\mathbf{v}_E$ is initialized using a ramped velocity profile:
\begin{equation}
    U_0 = \begin{cases}
        u_l, & x < -L/2 \\
        \frac{u_r-u_l}{L}(x+L/2)+u_l, & -L/2 \leq x \leq L/2\\
        u_r, & x > L/2
    \end{cases}, 
\end{equation}
with small-amplitude noise, it goes unstable. $L$ in this case is the total $x$-width of the ramped portion of the velocity profile. The evolution of the vorticity in this case is shown in Fig. \ref{fig:kh_contours}.

It is also useful to compare results produced by \hermes{} with theoretical linear growth rates. We begin by assuming a potential perturbation of the form $\varphi = \varphi_0(x) + \delta\phi(x)\exp (ik_zz-i\omega t)$ with a background velocity $U_0 = \varphi_0'/B \hat{z}$, and obtaining Rayleigh's equation,
\begin{equation}
    (\partial_x^2 - k_z^2)\delta \varphi = \frac{-k_z U_0^{''}}{\omega - k_z U_0}\delta\varphi.
    \label{eq:rayleigh}
\end{equation}

\begin{figure}
  \centering
  \includegraphics[]{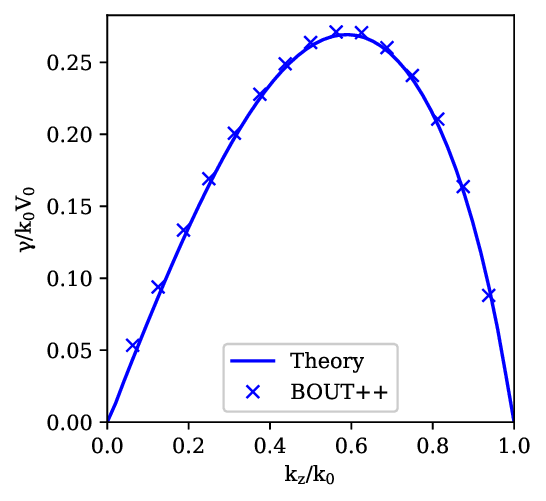}
  \caption{Linear growth rates for sinusoidal velocity profile with periodic boundary conditions. The estimated growth rates from \hermes{} closely match the curve produced by the semi-analytic model. }
\label{fig:kh}
\end{figure}

\noindent For a given velocity profile $U_0$, Eq. \eqref{eq:rayleigh} can be solved numerically to obtain a semi-analytic dispersion relation \citep{Zhu2018,Walkden2019}. This numerical solution is attained by discretizing the differential operator using finite differences and then computing the eigenvalues of the resulting linear system. That is, for a given wavenumber $k$, we solve the generalized eigenvalue problem for $\omega$:
\begin{equation}
    (A - k^2)(\omega -kU_0) = -kU_0'',
\end{equation}

\noindent where $A$ is the finite difference matrix. Following \cite{Zhu2018}, we apply a sinusoidal profile $U_0 = V_0\cos{(k_0x)}\hat{z}$ with periodic boundary conditions in both directions. In this case, $k_0 = 2\pi/L$, where $L$ is the size of the domain in the $x-$ and $z-$ directions. Linear growth rates from the simulation are obtained by fitting an exponential time dependence to the Fourier transform of the output of \hermes{} during the early, linear, phase of the simulation. Fig. \ref{fig:kh} shows a good agreement between \hermes{} result and the theoretical prediction. 

\subsection{Interchange instabilities \label{sec:interchange}}

The interchange instability is conceptually similar to the hydrodynamic Rayleigh-Taylor instability where a heavier fluid rests on top of a lighter fluid against the force of gravity. In a magnetized plasma, flute oscillations, so-called for their highly-anisotropic wavelengths ($k_\parallel \ll k_\perp$), are destabilized by charge separation arising from the discrepancy between the drift velocities of ions and electrons. The resulting azimuthal electric field amplifies the perturbation via a radial $\mathbf{E} \times \mathbf{B}$ drift \citep{Timofeev1992}. As the RDI instability is inextricably linked to flow shear and requires cylindrical coordinate, we defer its detailed analysis to \cref{sec:shear}.

To verify \hermes{}'s capability of capturing the interchange instability, a simple test case is set up in an isothermal plasma in a slab geometry. The diamagnetic current \eqref{eq:diamagnetic} is ultimately the term responsible for the interchange instability. The dominant contribution to the $\nabla \times \mathbf{b}/B$ term comes from the magnetic curvature, which can be approximated by $1/RB$, where $R$ is the radius of magentic field-line curvature. Consequently, the diamagnetic term reduces to $(1/R^2)\partial p/\partial\theta$, yielding an effective gravity proportional to $R^{-2}$. A constant background curvature was specified to simulate the effect in the slab geometry. 

\begin{figure}
    \centering
    \includegraphics[width=\linewidth]{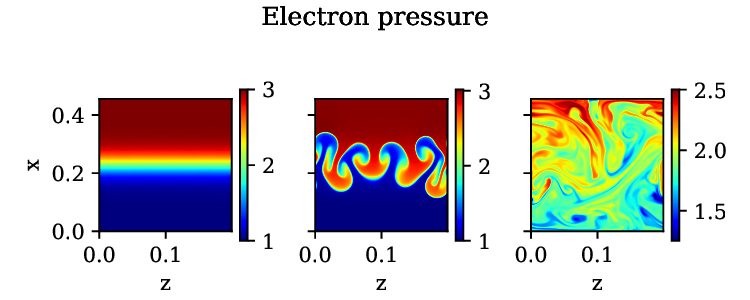}
    \caption{Time slices of electron pressure demonstrating interchange instability driven by magnetic curvature. The heavier fluid at the top eventually begins to mix with the lighter fluid at the bottom. }
    \label{fig:curv-interchange}
\end{figure}

The evolution of this curvature-driven interchange instability with \hermes{} is illustrated in Fig. \ref{fig:curv-interchange}. The initially laminar system quickly becomes unstable, exhibiting the ``mushroom structure'' characteristic of the interchange before transitioning into a fully mixed, turbulent state. A corresponding analysis for a non-isothermal plasma is presented in \cref{sec:FLR}, governed by the dispersion relation given in \eqref{eq:flr_dispersion}.

\subsection{Finite Larmor-radius effects \label{sec:FLR}}

A further study in \cite{Ng2005} investigated whether the finite Larmor-radius effects influence stability in the presence of a sheared flow. The finite spatial extent of particle gyration orbits in a magnetic field effectively smears the wave-particle resonance over a region on the scale of $\rho_i$. In fluid models, this kinetic effect can be approximated by incorporating gyroviscous terms.

The gyroviscous terms within the drift-reduced equations are not immediately apparent and thus merits brief discussion. \cite{Scott2007} demonstrated the correspondence between FLR effects in gyrofluid equations and the polarization and gyroviscous terms in long-wavelength ($k_\perp \rho_i \ll 1$) fluid equations. The lowest order of these effects are naturally accounted for in the vorticity equation \citep{Madsen2017}. The manifestation of FLR effects in the vorticity equation \eqref{eq:vorticity} appears two ways, first in the ``gyroviscous cancellation'', which cancels the advection of vorticity by the diamagnetic drift \citep{Simakov2003}, and second in the diamagnetic term in the vorticity.

To verify that \texttt{hermes-3} captures these lowest-order FLR effects, a series of simulations are carried out where the total interchange drive term $\nabla p_i + \nabla p_e$ is held constant, but the relative magnitudes of the ion and electron pressure gradients are swapped. An exponential background profile is used to maintain a constant gradient scale length across the simulation domain. We compare these numerical results against the theoretical dispersion relation for a CDI in a non-isothermal plasma (\citep{Madsen2017}):
\begin{equation}
    \lambda = \frac{\frac{4}{3}- \Bar{\kappa}_i \pm \sqrt{\left(\Bar{\kappa}_i-\frac{4}{3}\right)^2 - 4 k_\perp^{-2}\left(\Bar{\kappa}_i + \Bar{\kappa}_e -\frac{4}{3}\right)}}{2}.
    \label{eq:flr_dispersion}
\end{equation}

\noindent In this expression, $\lambda =\omega/k_z \xi$, $\bar{\kappa}_s = \nabla p_s/p_0\xi$, and $\xi = \rho_i/R$ is the normalized out-of-plane local radius of curvature.

\begin{figure}
    \centering
    \includegraphics[]{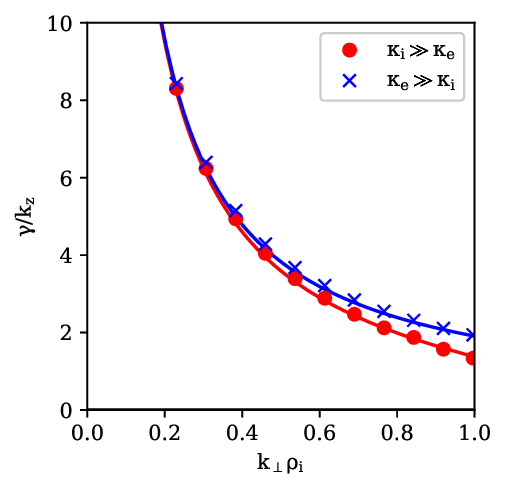}
    \caption{Interchange growth rates with ion and electron pressure gradients swapped. The larger gradient is a factor of 5 larger than the smaller gradient in both cases. For $\kappa_i \gg \kappa_e$, FLR effects are more prominent, causing the growth rate to decrease more rapidly as $k_\perp \rho_i \xrightarrow{} 1$. The \hermes{} simulation results (markers) show good agreement with the theoretical expression of \eqref{eq:flr_dispersion} (solid lines).}
    \label{fig:flr-effects}
\end{figure}

The simulation results, i.e., Fig. \ref{fig:flr-effects}, demonstrate a clear stabilizing effect for the case of a higher $\kappa_i$ as $k_\perp \rho_i \xrightarrow{}1$, showing excellent agreement with Eq. \eqref{eq:flr_dispersion}. In term of shear-flow stabilization, this gyroviscous effect increases the length scale at which turbulent eddies lose coherence by damping high-$k_\perp$, acting similarly to an enhanced collisional viscosity. As noted by \cite{Ng2005}, this stabilization effect is of roughly the same order of magnitude as collisional viscosity under reactor-relevant conditions. We postpone the discussion of FLR effects for RDI to \cref{sec:shear_flr} as it requires the relaxation of the Boussinesq approximation.

\section{Shear flow stabilization of interchanges\label{sec:shear}}

The interchange instability can be stabilized by strong flow shear. The primary stabilization mechanism is the tilting and stretching of turbulent eddies until their radial extent becomes small enough to be dissipated by collisional (viscous) or drift (gyroviscous) processes. A stability criterion derived by \cite{Hassam1992} is 
\begin{equation}
    V' > \gamma_g [\ln R_\mu^{1/3}]^{1/2} \sim \gamma_g,
    \label{eq:stability}
\end{equation}

\noindent where $\gamma_g \sim \left(g/L \right)^{1/2}$ is related to the linear growth rate, and $R_\mu = \gamma_g/\mu k^2$ is a Reynolds number. Because this criterion depends only weakly on the Reynolds number, we assume that $V' > \gamma_g$ is a sufficient condition for stabilization. This follows from the so-called ``quench rule'', explored first in \cite{Waltz1998} for a toroidal system. The effective gravity terms are approximately $g \sim c_s^2/R$ for CDI and $g\sim \nabla (v_\theta^2/2)$ for RDI. To study this stabilizing effect in \texttt{hermes-3}, we begin with a simplified isothermal model to isolate the CDI. 

\subsection{Shear stabilization of the CDI \label{sec:shear_stablization}}

As a test case, a parabolic, KH-stable, velocity profile is added to the interchange simulation, along with a linear pressure profile and constant viscosity. Initially, we set $T_i =0$ (the cold-ion approximation) to study the system without FLR effects, an assumption we relax later in this section. The interchange instabilities grow until they encountered the point where the velocity shear is sufficiently strong to completely stabilize them. The ansatz $V' > \gamma_g$ is used to predict this point, where we solve for the radial location $x_s$ beyond which the shear prevents the growth of the interchanges. Assuming a linear shear profile $V' = ax +v'_0$ and a linear pressure profile $p = bx+p_0$, $x_s$ is simply given by
\begin{equation}
    x_s = \frac{\sqrt{g/b}-v'_0}{a}. \label{eq:xs}
\end{equation}

\begin{figure}
    \centering
    \includegraphics[width=\linewidth]{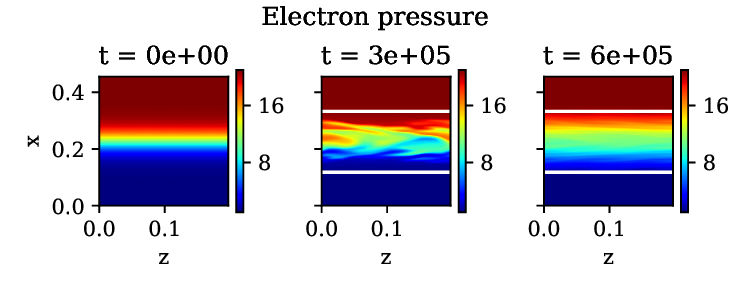}
    \caption{Interchange instability at $t=0$, the early stage, and later stage after being returned to a semi-laminar state by shear flow. The instability initially grows in the central region where the shear flow is weakest, before being sheared away as it spreads toward the edges. The white lines indicate the predicted width of the flattened region.}
    \label{fig:shear_stabilization}
\end{figure}

\noindent Fig. \ref{fig:shear_stabilization} shows the simulated CDI dynamics in the presence of flow shear; we observe that the analytical expression \eqref{eq:xs} approximates the width of the turbulent region quite well.

We then run simulations including FLR effects, again swapping the ion and electron pressure gradients to highlight the impact. We introduce source terms 
\begin{equation}
    S(u) = A(u_0 - u),
    \label{eq:source}
\end{equation}

\noindent to relax the profiles back toward a target state $u_0$, where $A$ as the relaxation timescale. These sources are applied to the density equation and to the Laplacian $\gradperp^2\varphi$ in the vorticity equation. The relaxation coefficient is chosen to relax the profile over timescales long compared to the observed growth of the instability. To minimize the influence of the boundary conditions, a hyperbolic tangent ($\tanh$) pressure profile is used.

\begin{figure}
    \centering
    \includegraphics[]{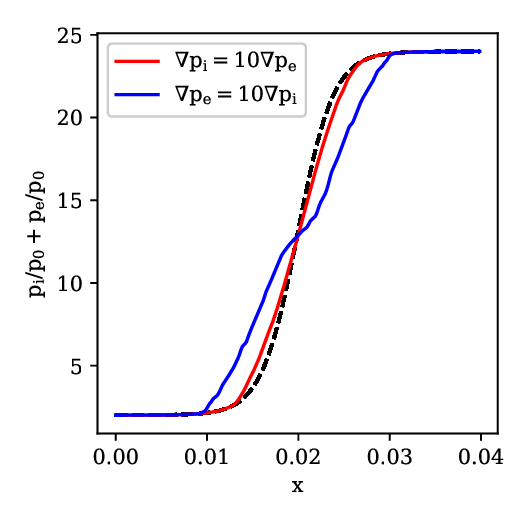}
    \caption{Interchange instability including FLR effects for a $\tanh$ initial profile. In the saturated state, the $z$-averaged total pressure profile for the case with $\nabla p_e \gg \nabla p_i$ is significantly more flattened. The relative pressure gradients in the high-FLR case are a factor of 10 larger than the low-FLR case. Here, $\rho_i/a \approx 0.1$, exaggerating the FLR stabilizing effect. The initial profile is shown with the dashed black line.}
    \label{fig:shear_compare}
\end{figure}

The results are shown in Fig. \ref{fig:shear_compare}, demonstrating that gyroviscosity can indeed play a significant role in stabilizing the interchange. The flattening of the total pressure profile in the low-FLR case is much more pronounced than that in the high-FLR case. For these simulations, the ratio of the Larmor radius to the domain size is $\rho_i/a >0.1$, which makes the stabilizing effect of a finite Larmor-radius more pronounced.

\subsection{Rotation-driven interchanges \label{sec:type2}}

We now examine the rotation-driven system. This instability is inherently more challenging to characterize because the azimuthal flow is simultaneously responsible for both the linear instability and nonlinear stabilization. We first employ a local model to obtain a linear growth rate that we will then compare to the flow shear in a manner similar to the previous section. Here we only consider peaked density and velocity profiles. Following \cite{Huang2004}, the density profile is chosen to be
\begin{equation}
    n = 1 + \frac{n_0}{2}\left[1-\cos\left(2\pi\left(x-D\left(x-x^2\right)\right)\right)\right],
\end{equation}

\noindent where $x = (R-R_i)/(R_o-R_i)$ and $D = (x_0-0.5)/(x_0(1-x_0))$. The parameter $D$ is used to shift the radial location of the peak of the density profile by picking $x_0 \in [0,1]$. The corresponding velocity profile is 
\begin{equation}
    v_{\theta,p} = v_0(R-R_o)(R-R_i).
    \label{eq:vthetap}
\end{equation}

In order to properly describe the RDI, we must relax the Boussinesq approximation. To see why this is necessary, we recast the vorticity equation from its original form in \cite{Simakov2003} to explicitly show the inertial term responsible for RDI, as derived in \cref{appB}. Using this formulation, Eq. \eqref{eq:vorticity} becomes 
  \begin{align}
       \pd{\omega}{t} &=  {\color{blue}-\nabla\cdot(\omega\ve)} \nonumber \\ &
     {\color{red}-\nabla\cdot\left[ 
 \frac{m_i}{2B^2} \gradperp(\ve\cdot\nabla p_i) +\frac{n}{2B^2} (\gradperp^2\varphi) \mathbf{v}_{di} -  \nabla\cdot \left(\frac{m_i}{2B^2} \gradperp p_i\right)\ve
       \right]} \nonumber \\ &
       \color{ForestGreen} +\nabla \cdot \left[ 
\frac{m_i(\ve \cdot \ve)}{2} \frac{\mathbf{b}\times\nabla n}{B}   
       \right] \nonumber  \\& + \nabla\cdot(\mathbf{J}_d + \mathbf{J}_{ci}), \label{eq:vorticity_full}
\end{align}  

\noindent with
\begin{equation}
    \mathbf{v}_{di} = \frac{\mathbf{b}\times \nabla p_i}{nB},
\end{equation}

\noindent and 
\begin{equation}
    \omega = \nabla \cdot \left[\frac{m_i}{B^2}(n\gradperp\varphi+\gradperp p_i)\right].
\end{equation}

\begin{figure}
    \centering
    \includegraphics[width=\linewidth]{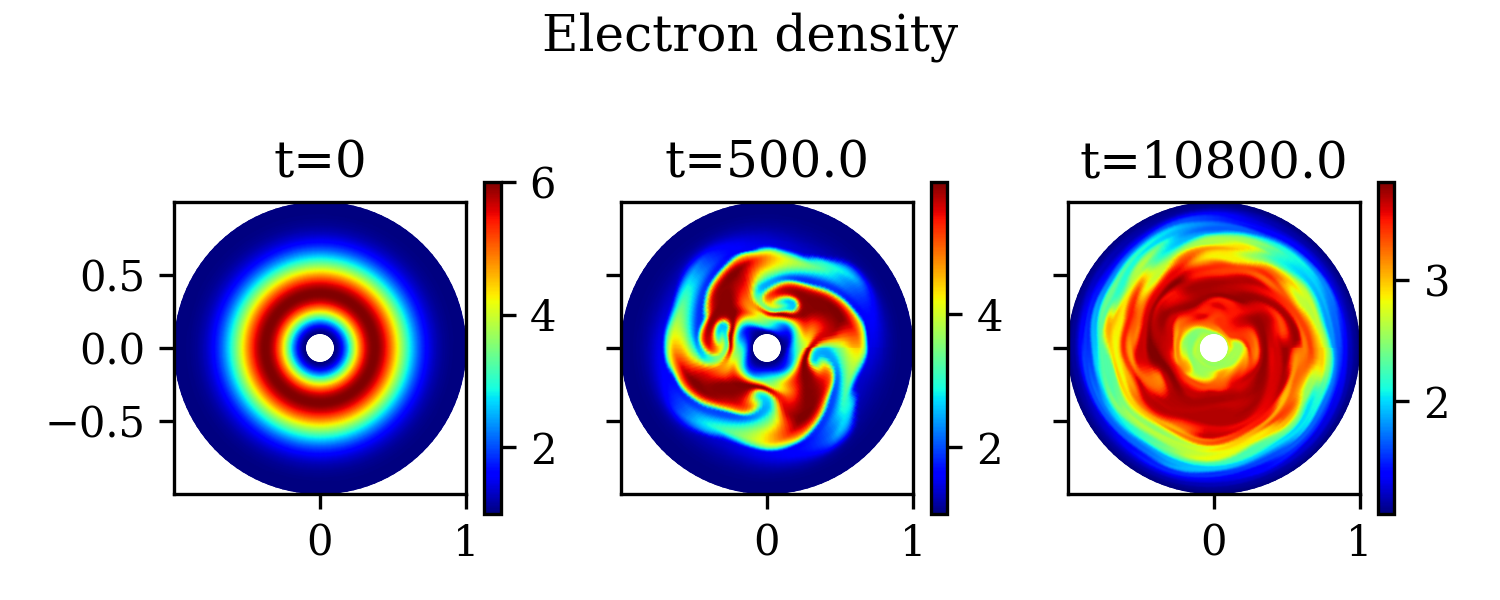}
    \caption{Evolution of a centrifugally driven interchange in a M=4 background flow. The strong centrifugal force due to azimuthal rotation drives instability. The flow shear eventually recreates a stratified density profile with a more gradual gradient. For this case with a small relaxation source, the flattening is dramatic, and vortices eventually appear near the edge in the long-time state.}
    \label{fig:cent_instability}
\end{figure}

\noindent Writing the vorticity equation in the form \eqref{eq:vorticity_full} is beneficial because it separates the polarization current into the {\color{blue} advection of vorticity} (first line), {\color{red} diamagnetic effects} (second line), and the {\color{ForestGreen} plasma inertia} (third line). This recast form was also found to exhibit better numerical performance when implemented in \hermes{}. The inertial term is ultimately responsible for exciting the RDI, and is similar to the Equation (22) in \cite{Angus2014}, which also explored the removal the Boussinesq approximation. This form of the vorticity equation is similar to that found in \cite{Popovich2010}, which studied inertia-driven interchanges and Kelvin-Helmholtz modes in the linear phase. This paper also derived an expression similar to \eqref{eq:appB:growth_rate} for the linear growth rate of the interchange for the specific case of the a uniform rotation profile. An example of this centrifugally driven interchange is shown in Fig. \ref{fig:cent_instability}. Note that these simulations assume an isothermal plasma, an assumption that will be relaxed in Section \cref{sec:global}. All simulations are performed on a $256\times256$ grid in the $R-\theta$ plane with Neumann boundary conditions in the $R-$direction and periodic boundary conditions in the $\theta-$direction. 

\subsubsection{Stability criterion \label{sec:type2stability}}

\begin{figure}
    \centering
    \includegraphics[width=\linewidth]{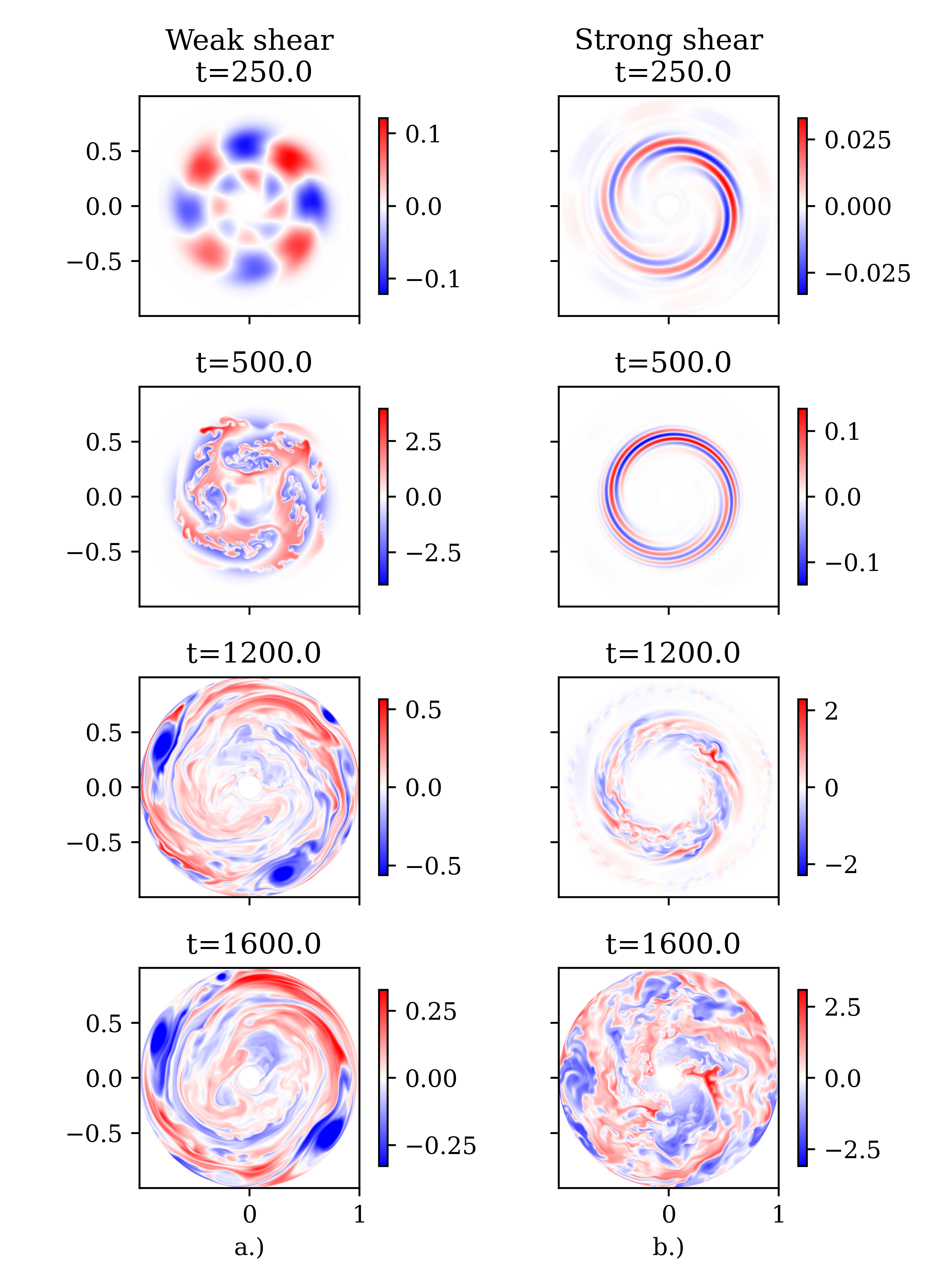}
    \caption{Perturbed density $(n-\langle n\rangle_\theta)/N_0$ for the weak (a) and strong (b) shear cases. The weak shear case exhibits growing modes largely unaffected by shear, while in the strong shear case the modes are highly sheared from the onset. Both cases eventually develop into turbulence. Times are normalized to the ion cyclotron frequency.}
    \label{fig:shear_regimes_modes}
\end{figure}

We cannot construct a simple expression similar to Eq. \eqref{eq:xs} for the RDI because the azimuthal flow is responsible for both instability and shear stabilization. However, we can still compare the shearing rate to the linear growth rate,
\begin{equation}
    |V'| > \sqrt{-\frac{(V^2)'/2}{L_n}}.
\end{equation}

\noindent The expression on the right represents the inertial force divided by the density length scale. A full derivation of the growth rate is provided in \cref{appC}. Solving for $L_n = (d\ln n/dR)^{-1}$ yields 
\begin{equation}
    |L_v|/L_n > -1.
    \label{eq:criterion}
\end{equation}

This is quite similar to the Richardson number \citep{Terry2000}, which also compares the shearing rate with the linear growth rate of the interchange. If Eq. \eqref{eq:criterion} is satisfied everywhere in the domain, we posit that instabilities due to plasma inertia like the interchange will be shear stabilized. Note that this relationship is only valid locally, and cannot guarantee global stability. An example of such a global instability is the KH mode, which we test in Section \cref{sec:global}. Additionally, we differentiate between the so-called ``strong'' and ``weak'' regimes based on whether the density gradient is positive or negative at the radial location of the peak velocity, respectively. This categorization is similar to the two regimes studied in \cite{Ivanov2025} for ITG turbulence, which also compare the shearing rate with the growth rate of the underlying instability. The delineation between strong and weak shear also follows the observation in \cite{Huang2004} of how the sign of $dn/dR$ at the velocity peak location alter the system behavior. However, in that paper it was concluded that the system would be shear-stabilized if $dn/dR > 0$ at the peak of the velocity; here we can only conclude that the system will experience strong shear. There are a number of possible reasons for this discrepancy, but it is likely because the stable profile used in \cite{Huang2004} which satisfied Eq. \eqref{eq:criterion} everywhere.

\begin{figure}
    \centering
    \includegraphics[width=\linewidth]{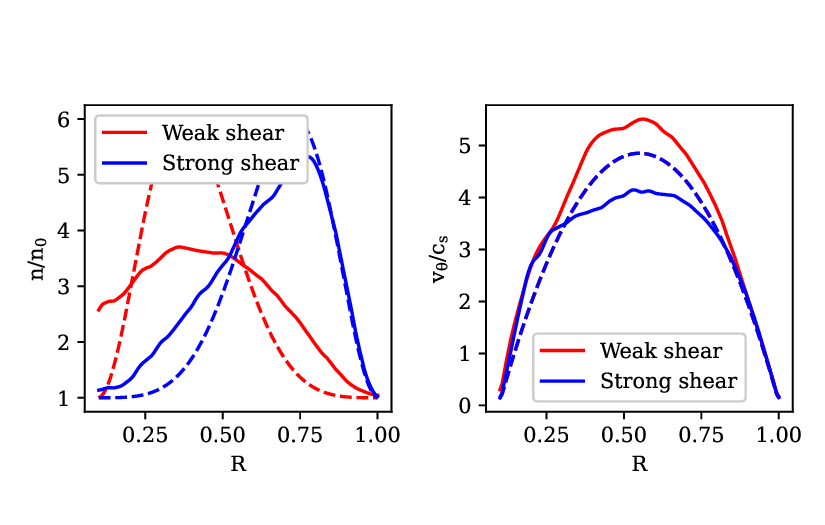}
    \caption{Evolution of mean profiles for the strong and weak shear cases. The degree of density flattening between initial (dashed) and quasi-steady-state (solid) profiles is substantially more intense in the weak shear case.}
    \label{fig:cent_sw}
\end{figure}

Fig. \ref{fig:shear_regimes_modes} illustrates the mode evolution of these different regimes. The strong shear regime exhibits highly sheared eddies concentrated in the middle where the flow shear is smallest, whereas the radial wave numbers of the weakly sheared modes are largely unaffected. The shear also significantly delays the onset of fully developed turbulence, especially in the strong shear case. 
Fig. \ref{fig:cent_sw} shows the evolution of the mean density and velocity profiles. The density profile is flattened to a much more greater extent in the weak shear case than in the strong shear case, though both cases experience some degree of gradient relaxation.

\begin{figure}
    \centering
    \includegraphics[]{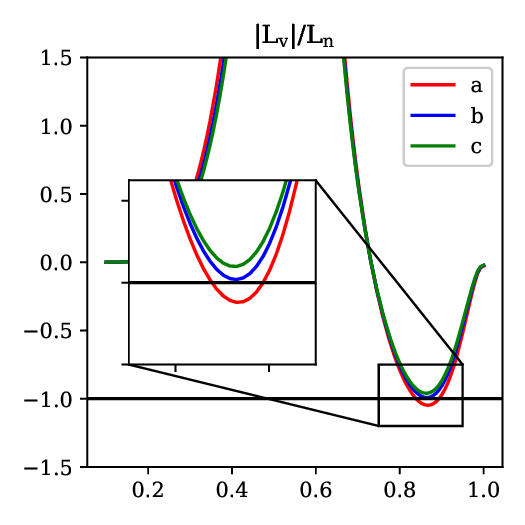}
    \caption{Evaluation of the stability criterion Eq. \eqref{eq:criterion} for three slightly different profiles (see figure \ref{fig:cent_stability}). These profiles were carefully selected to lie very close to the stability threshold, denoted by the black line at $y=-1$.}
    \label{fig:criterion}
\end{figure}

\begin{figure}
    \centering
    \includegraphics[width=\linewidth]{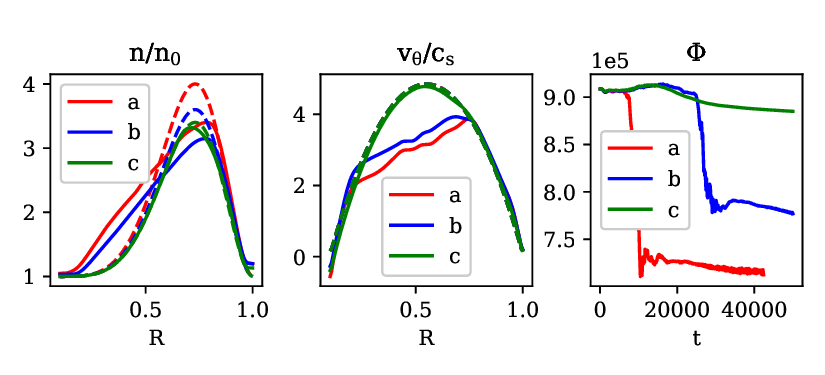}
    \caption{Profile evolutions of three simulations with slightly different density peaks, demonstrating the sensitivity of the system to the stability criterion. Cases a) and b) experience turbulence that crashes the potential while case c) varies smoothly.}
    \label{fig:cent_stability}
\end{figure}

In cases where Eq. \eqref{eq:criterion} is satisfied everywhere, we find that the interchange does indeed appear to be stabilized by shear. The system's temporal evolution is highly sensitive to this criterion, with slight deviations producing dramatic differences. Fig. \ref{fig:cent_stability} shows the evolution of three slightly different profiles and the total potential drop across the plasma, while Fig. \ref{fig:criterion} plots Eq. \eqref{eq:criterion} for these cases. These simulations were performed with the relaxation sources turned off to test for full shear stabilization. 
In all three cases, fluctuations always originate in the regions where the shear is weakest. In the unstable and marginal unstable cases, these fluctuations eventually propagate inward to the region where the profile is linearly unstable to RDI, triggering the characteristic turbulence of the interchange instability. In contrast, the stable run experiences slight profile evolution but never devolves into full turbulence over the course of the simulation, as evidenced by the steadiness of the potential drop whereas the unstable cases experience a dramatic crash in the potential, marking the onset of turbulence. This sensitivity could be an important design consideration for target profiles in a rotating mirror machine.

While this section outlines broad qualitative observations, the quantitative behavior of these systems is highly nonlinear and profile-dependent. Particularly for this paper, we are interested in how the profile evolution is affected by KH-modes. We highlight this interaction in \cref{sec:global}, where we compare a quartic (i.e., KH-unstable) and parabolic (i.e., KH-stable) velocity profile.  

\subsubsection{Finite Larmor-radius effects in RDI\label{sec:shear_flr}}

To isolate FLR effects in the RDI cases, we artificially remove the diamagnetic terms from the vorticity equation. That is, the terms in red in Eq. \eqref{eq:vorticity_full} are omitted, and the vorticity becomes 
\begin{equation}
    \omega = \nabla\cdot\left(\frac{n m_i}{B^2}\gradperp\varphi\right).
\end{equation}

\begin{figure}
    \centering
    \includegraphics[width=\linewidth]{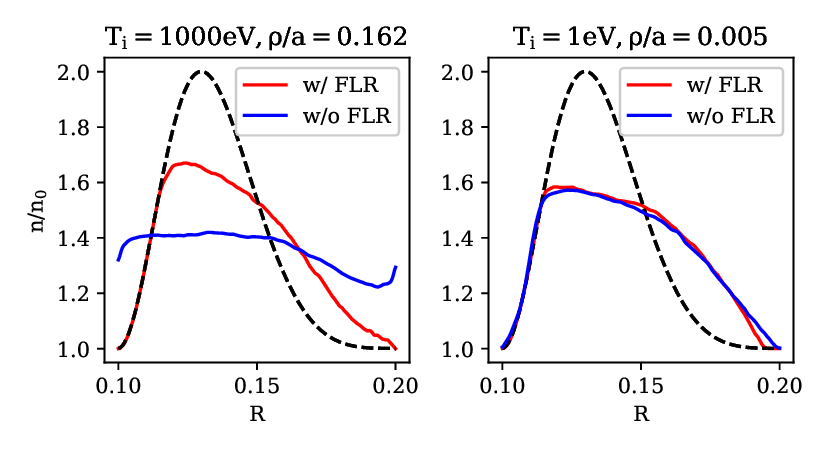}
    \caption{Steady state profiles with and without FLR stabilization for two different values of ion temperature. The corresponding value of $\rho/a$ is also shown. The velocity (not shown) was the same in both cases. Large relaxation sources for the velocity (10-100x those used for other simulations) were needed to maintain consistency in the azimuthal velocity, but this deemed to be the best way to isolate the FLR comparison. No relaxation source was used for the density. The figure on the left has a large ion temperature, and the figure on the right has a small ion temperature. As expected, the effects of the finite Larmor-radius are much more pronounced in the case of larger ion temperatures.}
    \label{fig:cent_flr}
\end{figure}

\noindent The FLR stabilization ultimately comes from the diamagnetic term, as shown in the first term under the square root in the dispersion relation for CDI, Eq. \eqref{eq:flr_dispersion}. During these tests, we also artificially pin the velocity to the initial condition to highlight the FLR effects on the density profile evolution. 
Fig. \ref{fig:cent_flr} demonstrates this effect. The simulation was run with both large and small ion temperatures to vary $\rho_i/a$. As expected, the FLR effects are much more pronounced in the large ion temperature case. Ultimately, because interchanges are intrinsically long-wavelength modes, the effect of FLR stabilization is minimal unless $\rho_i/a$ is large, which is undesirable for confinement reasons in a fusion reactor. This is evident in Fig. \ref{fig:flr-effects}, which shows the reduction in linear growth rate only becomes significant as $k_\perp \rho_i \xrightarrow[]{} 1$. It is also worth noting that the inclusion of additional physics terms could alter the system's macroscopic behavior, making it difficult to cleanly isolate FLR effects in fully turbulent, unpinned simulations.

\section{Kelvin-Helmholtz-interchange coupled instability\label{sec:global}}
\begin{figure}
    \centering
    \includegraphics[width=\linewidth]{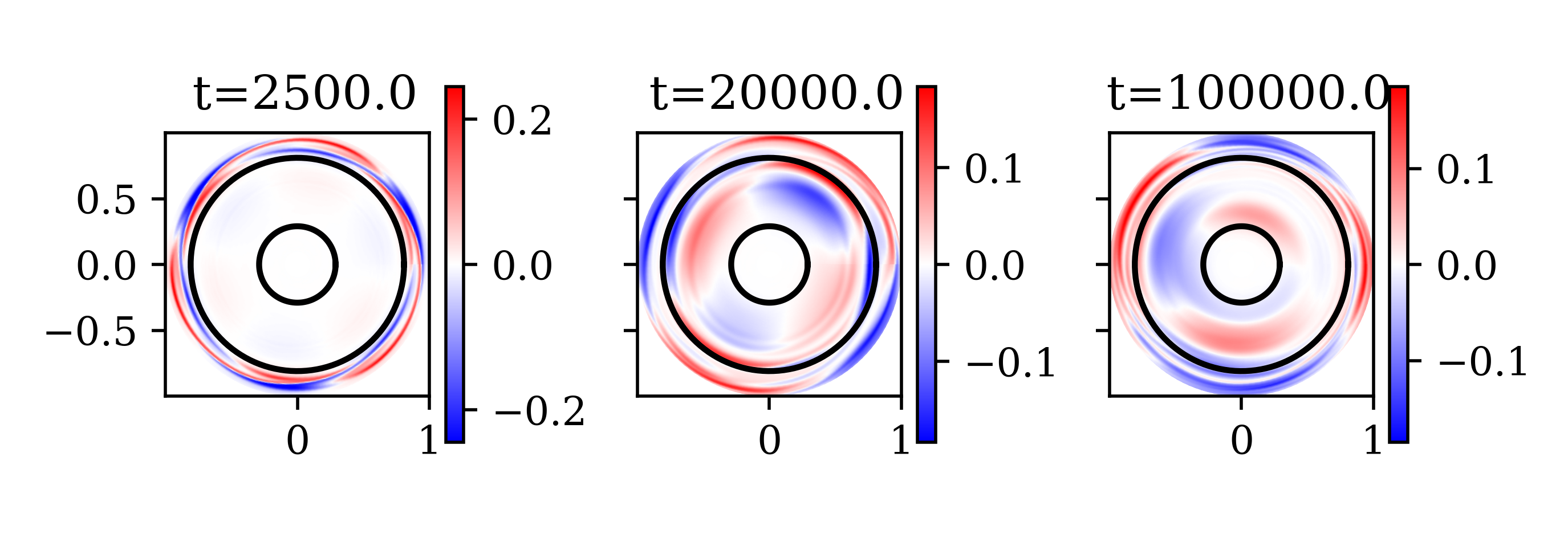}
    \caption{Mode structure of perturbed density $(n-\langle n\rangle_\theta)/N_0$ for quartic profiles in the stable RDI regime. The quartic profile exhibits low frequency $k_\theta$ modes throughout the plasma. The black circles in the quartic profile mark the inflection points.}
    \label{fig:global_modes_stable}
\end{figure}

\begin{figure}
    \centering
    \includegraphics[width=\linewidth]{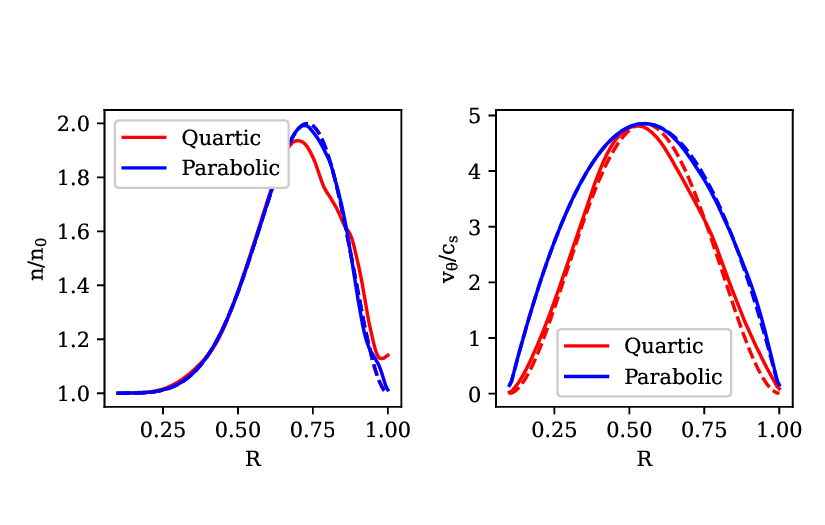}
    \caption{Steady state profiles for quartic and parabolic velocity profiles. }
    \label{fig:global_profiles_stable}
\end{figure}
 In this section we examine velocity profiles that are not inherently KH-stable. The inclusion of inflection points into the velocity profile results in unstable modes that are potentially small in amplitude by themselves but may drive unstable interchange modes. We now explore this possibility by comparing the behavior of parabolic (KH-stable) and quartic (KH-unstable) velocity profiles. The quartic profile is $v_{\theta,q} = (v_{\theta,p})^2/v_0$, with $v_{\theta,p}$ given by Eq. \eqref{eq:vthetap}. The quartic velocity possesses inflection points at $R = (R_i+R_o(2 \mp\sqrt3))/(3\mp\sqrt{3})$, corresponding to roughly 25\% and 75\% of the radial domain. These simulations are performed in the $R-\theta$ plane using $256\times256$ resolution.

We first consider profiles well below the RDI stability threshold, as defined by Eq. \eqref{eq:criterion}. The profile with inflection points exhibits global fluctuations that grow to a finite amplitude and then saturate, as shown in Fig. \ref{fig:global_modes_stable}. These fluctuations peak around the inflection points, suggesting a KH-type mode. The steady-state profiles are shown in Fig. \ref{fig:global_profiles_stable}. Although the quartic velocity profile undergoes more modification, the overall difference is minimal, suggesting that the KH mode only is not catastrophic.

\begin{figure}
    \centering
    \includegraphics[width=\linewidth]{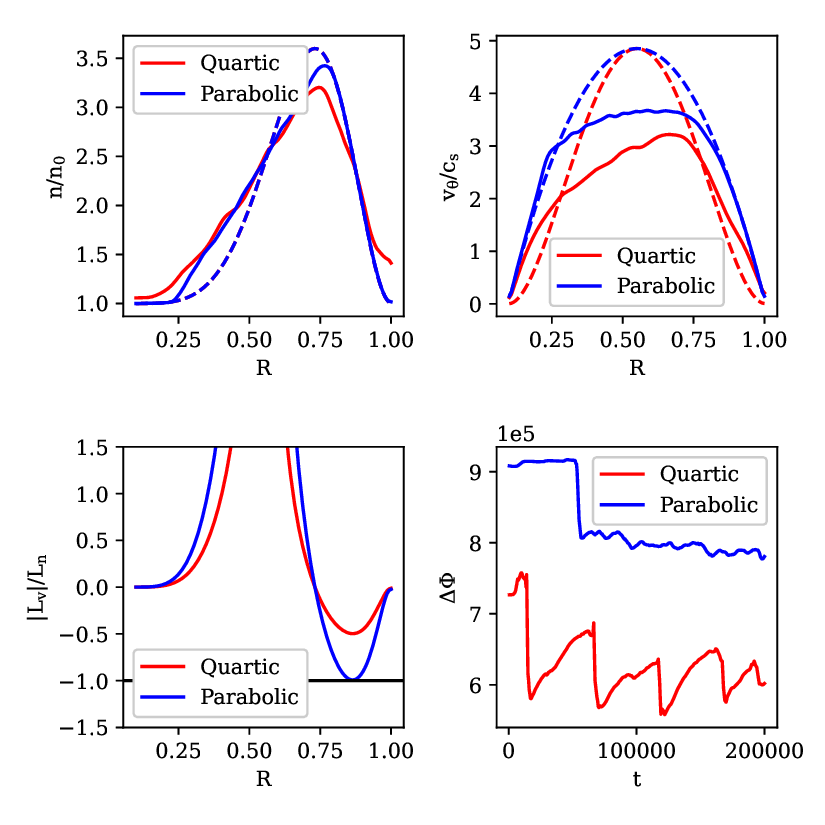}
    \caption{Steady state profiles for quartic (a) and parabolic (b) velocity profiles. The steady state density profiles are similar for both cases, but the steady state peak of the velocity is lower. Case (a) goes unstable even though \eqref{eq:criterion} is satisfied everywhere, indicating that the KH mode is possibly driving the instability. The behavior of the potential drop across the plasma is also different, as case (a) shows a cyclical behavior in the potential crashes, while the potential crashes only once in case (b).}
    \label{fig:global_profiles_unstable}
\end{figure}

Next, we compare a case in the strong shear regime with a quartic velocity profile possessing the same peak velocity. The steady state profiles along with the stability criterion evaluated via Eq. \eqref{eq:criterion} are shown in Fig. \ref{fig:global_profiles_unstable}. Although the quartic case is initially well below the instability threshold, the resulting KH mode was able to trigger interchanges. Furthermore, the quartic case experiences cyclical crashes in the radial potential drop. While the steady-state density profiles remain similar between the two cases, the KH-unstable profile exhibits enhanced instability near the radial boundaries.

\section{Conclusion}

In this study, we performed numerical experiments investigating the global stability of rapidly rotating plasmas to Kelvin-Helmholtz (KH) and interchange modes, connecting our observations with theoretical predictions. We examined shear-flow stabilization in the context of curvature-driven interchange (CDI), validating a comparison between the shearing rate and the linear growth rate. Additionally, the drift-reduced fluid equations were modified to relax the Boussinesq approximation, allowing us to explicitly resolve the inertial term responsible for the rotation-driven interchange (RDI). The interplay between RDI and shear-flow stabilization proved to be of crucial importance in determining the long-time evolution of the initial laminar profiles. 

We identified two distinct behavioral regimes for the RDI (so-called ``weak'' and ``strong'') and proposed a localized criterion to predict the susceptibility of a given profile to this instability. In the case within the strong-shear regime that \eqref{eq:criterion} is satisfied everywhere, we identify a stable regime that won't experience turbulent profile crashes. We also analyzed the role of finite Larmor-radius (FLR) effects in the stabilization of these interchange modes.
Finally, we observed how KH-unstable velocity profiles affect the global stability of the system, and concluded that the impact can be significant when the system resides close to the stability boundary. Consequently, we advise caution when relying solely on Eq. \eqref{eq:criterion} to predict the stability of a profile, as other instabilities present in the system could eventually drive RDI even if the initial configuration appears to be fully shear-stabilized. Nevertheless, our observations clearly indicate that the weak shear regime should be avoided whenever possible, as it produces the most intense turbulence.

Future work will focus on extending this azimuthal model to a fully 3D mirror geometry, which is expected to alter the dynamics by suppressing the $k_\parallel = 0$ mode. 

\section{Acknowledgments}

Research support for this work was provided by the U.S. Department of Energy. E. Tocco and I. G. Abel was supported by contract number DESC0024425. I. G. Abel was also supported by DOE contract DEFG0293ER54197. Prepared in part by Lawrence Livermore National Laboratory under contract DE-AC52-07NA27344 and LLNL-JRNL-2015332. This research used resources of the National Energy Research Scientific Computing Center (NERSC), a Department of Energy User Facility using NERSC award FES-ERCAP0036156. 

\appendix

\section{}\label{appB}
In attempting to perform simulations that explore RDI, it was found that the vorticity equation in its original form from \cite{Simakov2003} behaved unexpectedly when trying to isolate the RDI. It is for this reason that we endeavor to rewrite it here to improve readability and performance in studying the RDI. We start with equation (D3) for the polarization current in \cite{Simakov2003} and, only considering the electrostatic part, 

\begin{equation}
    qn\mathbf{v}_{pi} = qn\frac{\hat{b}}{\Omega_i} \times \left(\pd{}{t} + \mathbf{v}_E \cdot \nabla\right)\mathbf{v}_E \approx -\left(\pd{}{t} + \mathbf{v}_E \cdot \nabla\right)\left[\frac{nm_i}{B^2}\gradperp \varphi \right].
\end{equation}

A vector identity is used to rewrite the convective term

\begin{equation}
    \ve\cdot \nabla \left(\frac{nm_i}{B^2}\gradperp\varphi\right) \approx \frac{1}{2}\left[\left(\frac{m_i}{B^2} \ve \cdot \nabla n\right)\gradperp\varphi + \omega \ve + \left(\frac{n m_i}{B^2}\gradperp^2\varphi\right)\ve\right].
\end{equation}

By adding and subtracting $\frac{1}{2}\nabla \cdot \omega\ve$, and using $\nabla\cdot \frac{nm_i}{B^2}\gradperp\varphi = \frac{m_i}{B^2} \gradperp\varphi \cdot \nabla n + \frac{n m_i}{B^2}\gradperp^2 \varphi$, we obtain

\begin{equation}
\begin{split}
    \ve\cdot \nabla \left(\frac{nm_i}{B^2}\gradperp\varphi\right) \approx \omega \ve + \frac{1}{2}\left(\frac{m_i}{B^2} \ve \cdot \nabla n\right)\gradperp\varphi - \frac{1}{2}\frac{m_i}{B^2}(\gradperp\varphi \cdot \nabla n) \ve.
\end{split}
\end{equation}

Recognizing the last two terms on the right represent the inertial term and rewriting as a vector triple product 

\begin{equation}
qn \vc =     \frac{m_i}{2B^2}\left[\left(\ve \cdot \nabla n\right)\gradperp\varphi - \ve(\gradperp\varphi \cdot \nabla n) \right] = \frac{m_i}{2B^2} \nabla n \times(\ve \times \gradperp \varphi).
\end{equation}

Recognizing $\gradperp \varphi = -B\mathbf{b}\times \ve$, the term in parentheses can be expanded

\begin{equation}
    -\ve \times(\mathbf{b} \times \ve) = -\mathbf{b} (\ve\cdot\ve) + \ve(\mathbf{b}\cdot\ve) = -\mathbf{b}(\ve \cdot \ve),
\end{equation}

\noindent we then have

\begin{equation}
     -\frac{m_i}{2B} \nabla n \times \mathbf{b}(\ve\cdot\ve) = \frac{m_i}{B}\frac{\ve \cdot \ve}{2} \mathbf{b}\times\nabla n  \label{eq:cent_drift}.
\end{equation}

The expression for the polarization drift is then 

\begin{equation}
    \pd{\omega}{t} = -\nabla\cdot \omega \ve + \nabla \cdot \left[ \frac{m_i\ve \cdot \ve}{2} \frac{\mathbf{b}\times\nabla n}{B} \right].
\end{equation}

\section{}\label{appC}

In this section, we derive the linear growth rate of the RDI ignoring FLR and other effects. Expressions including other physics were obtained, but found to be unnecessary for our purposes and obscured the interpretation of the results. Isolating the terms responsible for RDI and assuming cold ions ($T_i=0$), we obtain expressions for continuity and charge conservation
\begin{equation}
    \pd{n}{t} = -\nabla \cdot (n \ve),
\end{equation}
and
\begin{equation}
    \pd{}{t}\left[\nabla\cdot\left(n\frac{m_i}{B^2}\gradperp\varphi\right)\right] = \nabla \cdot\left[\frac{m_i}{B}\frac{\ve \cdot \ve}{2} \mathbf{b}\times\nabla n \right].
\end{equation}

Linearizing about a background profile, the density and electrostatic potential become

\begin{equation}
    n = n_0 + \delta n \exp (-i\omega t+i\mathbf{k}\cdot\mathbf{x}),
\end{equation}
and
\begin{equation}
    \varphi = \varphi_0 + \delta \varphi \exp (-i\omega t+i\mathbf{k}\cdot\mathbf{x}).
\end{equation}

Because the fastest growing mode for interchanges is usually when $k_R=0$, we only consider $\mathbf{k} = \frac{m}{R}\hat{\phi}$.

The $\mathbf{E}\times\mathbf{B}$ velocity is
\begin{equation}
    \ve = \frac{\mathbf{b}\times\gradperp\varphi}{B} = \frac{1}{B}\left(-\hat{R}\delta\varphi\frac{im}{R}+\hat{\phi}\varphi_0'\right).
\end{equation}

Ignoring higher order terms, the density equation becomes
\begin{equation}
    -i\omega \delta n = -\nabla\cdot\left[(n_0+\delta n)\frac{1}{B}\left(-\hat{R}\delta\varphi\frac{im}{R}+\hat{\phi}\varphi_0'\right)\right] = n_0'\delta\varphi\frac{im}{RB}-\delta n \varphi_0'\frac{im}{RB}.
\end{equation}
Solving for the perturbed density
\begin{equation}
    \delta n = -\frac{n_0'}{\omega-U_0\frac{m}{R}}\delta\varphi \frac{m}{R},
    \label{eq:appB:delta_n}
\end{equation}

\noindent where $U_0 \equiv \varphi_0'/B$. 

We now linearize the vorticity equation,
\begin{align}
    -i\omega  \left[n_0\frac{m_i}{B^2}\delta\varphi \frac{m^2}{R^2}\right] &= \nabla\cdot\left[\frac{m_i}{2B^2}\left(U_0^2 + U_0\delta\varphi\frac{im}{RB}\right)\left(-\hat{R}\delta n \frac{im}{R}+\hat{\phi}n_0'\right)\right] \\ &=
    \frac{i m}{R} \frac{m_i}{B^2} (U_0^2/2)'\delta n - U_0\frac{m^2}{R^2}\frac{m_i}{2B^2}n_0'\delta\phi. \nonumber
\end{align}
Simplifying and plugging in \eqref{eq:appB:delta_n} for $\delta n$
\begin{equation}
    \left(\omega-\frac{U_0}{2}\frac{n_0'}{n_0}\right)\delta\varphi = -(U_0^2/2)'\frac{n_0'/n_0}{\omega-U_0\frac{m}{R}}\delta\varphi,
\end{equation}
\noindent we then obtain a dispersion relation
\begin{equation}
     \left(\omega-\frac{U_0}{2}\frac{n_0'}{n_0}\right)\left( \omega - U_0\frac{m}{R}\right) = -(U_0^2/2)'\frac{n_0'}{n_0}.
     \label{eq:appB:dispersion_rln}
\end{equation}

Taking the maximum growth rate occurring at $m/R = (1/2)n_0'/n_0$, found by taking the derivative of \eqref{eq:appB:dispersion_rln} with respect to $m/R$, we can solve for $\omega$
\begin{equation}
    \omega = \frac{U_0}{4}\frac{n_0'}{n_0} \pm \sqrt{ -(U_0^2/2)'\frac{n_0'}{n_0}}.
 \end{equation}
\noindent For $\omega = \omega_r - i\gamma$, the imaginary part gives the expression for the linear growth rate
\begin{equation}
    \gamma^2 = -(U_0^2/2)'\frac{n_0'}{n_0}.
    \label{eq:appB:growth_rate}
\end{equation}

\bibliography{bibliography}
\bibliographystyle{jpp}

\end{document}